\documentclass[%
showkeys,
 amsmath,amssymb,
 aps,
]{revtex4-2}

\usepackage{dcolumn}
\usepackage{bm}
\usepackage{subfigure}
\usepackage{subfigmat}
\usepackage[dvips]{dropping}
\usepackage{hyperref}
\hypersetup{
    colorlinks = true,
    linkcolor = {blue},
    citecolor  = {blue},
    filecolor = {blue},
    menucolor= {blue},
    runcolor = {blue},
    urlcolor = {blue},
    linkcolor = {blue}
}
\usepackage{breakurl}
\usepackage{psfrag}
\usepackage{latexsym}
\usepackage{amsmath}
\usepackage{multirow}
\usepackage{amssymb}
\usepackage{epsfig}
\usepackage{graphicx}
\usepackage{floatrow}
\usepackage{units}
\usepackage{array}
\usepackage{xfrac}
\usepackage{bm}
\usepackage{bigints}
\usepackage{color}
\usepackage{tikz}
\usepackage{tikzscale}
\usepackage{pgfplots}
\usepackage{pifont}
\usepackage{mathtools}
\usetikzlibrary{decorations}
\usetikzlibrary{calc,decorations.pathreplacing}
\usepackage{flushend}
\usepackage{upgreek}
\usepackage{scrextend}
\usepackage{lipsum}
\usepackage{enumitem}

\newcommand*\HR[1]{$\mathcal{R}${#1}}
\newcommand*\LF{$\mathcal{LF}$}
\newcommand*\NF{$\mathcal{NF}$}
\newcommand*\HF{$\mathcal{HF}$}

\begin{document}


\title{Inner-scaled Helmholtz resonators with grazing turbulent boundary layer flow}

\author{Giulio Dacome}
\author{Renko Siebols}
\author{Woutijn J. Baars}\email{w.j.baars@tudelft.nl}
\affiliation{Faculty of Aerospace Engineering, Delft University of Technology, 2629 HS Delft, The Netherlands}


\date{\today}

\begin{abstract}
Response details are presented of small-scale Helmholtz resonators excited by grazing turbulent boundary layer flow. A particular focus lies on scaling of the resonance, in relation to the spatio-temporal characteristics of the near-wall velocity and wall-pressure fluctuations. Resonators are tuned to different portions of the inner-spectral peak of the boundary-layer wall-pressure spectrum, at a spatial scale of $\lambda_x^+ \approx 250$ (or temporal scale of $T^+ \approx 25$). Following this approach, small-scale resonators can be designed with neck-orifice diameters of minimum intrusiveness to the grazing flow. Here we inspect the TBL response by analysing velocity data obtained with hot-wire anemometry and particle image velocimetry measurements. This strategy follows the earlier work by \citet{panton:1975a} in which only the change in resonance frequency, due to the grazing flow turbulence, was examined. Single resonators are examined in a boundary layer flow at $Re_\tau \approx 2\,280$. Two neck-orifice diameters of $d^+ \approx 68$ and 102 are considered, and for each value of $d^+$ three different resonance frequencies are studied (targeting the spatial scale of $\lambda_x^+ \approx 250$, as well as sub- and super-wavelengths). Passive resonance only affects the streamwise velocity fluctuations in the region $y^+ \lesssim 25$, while the vertical velocity fluctuations are seen in a layer up to $y^+ \approx 100$. A narrow-band increase in streamwise turbulence kinetic energy at the resonance scale co-exists with a more than 20\,\% attenuation of lower-frequency (larger scale) energy. Current findings inspire further developments of passive surfaces that utilize the concept of changing the local wall-impedance for boundary-layer flow control, using miniature resonators as a meta-unit.
%
\end{abstract}

\maketitle

\section{Introduction}\label{sec:intro}
Turbulent boundary layer (TBL) flows comprise intense velocity and (wall-)pressure fluctuations, which are responsible for turbulence kinetic energy (TKE) production, energy transport and the generation of surface quantities such as friction and heat transfer. For flow control purposes, \emph{passive} methods are preferred based on practicality. Given that there is a growing body of knowledge suggesting that surface-embedded Helmholtz resonators (HRs) can substantially alter the kinetic energy content of grazing wall-bounded turbulence, our current work investigates how \emph{miniature} resonators can be used to alter the flow. Hereby we aim to answer how the scaling of spatio-temporal characteristics of a resonator (relative to the broadband range of scales of near-wall turbulence) affects the local change in the wall-impedance. Throughout this manuscript, a Cartesian coordinate system ($x$, $y$, $z$) is employed to denote the streamwise, wall-normal and spanwise directions of the flow, respectively. The corresponding quantities $u$, $v$, $w$ and $p$ represent the Reynolds-decomposed fluctuations of the velocity components and the pressure. The friction Reynolds number $Re_\tau \equiv \overline{U}_\tau \delta/\nu$ is the ratio of the inertial length scale (the boundary layer thickness, $\delta$) to the viscous length scale, $l^* = \nu/\overline{U}_\tau$, where $\nu$ is the kinematic viscosity and $\overline{U}_\tau = \sqrt{\tau_w/\rho}$ is the friction velocity (here $\tau_w$ is the wall-shear stress and $\rho$ is the fluid density). When a length is inner-scaled with the viscous scale $l^*$, when time is scaled with $\nu/\overline{U}_\tau^2$, and when a velocity is normalized with $\overline{U}_\tau$, the quantity is presented with a superscript `$+$'.

\subsection{Influencing the near-wall cycle of turbulent boundary layer flow}\label{sec:introcontrol}
\begin{figure*}[htb!] 
\vspace{0pt}
\centering
\includegraphics[width = 0.999\textwidth]{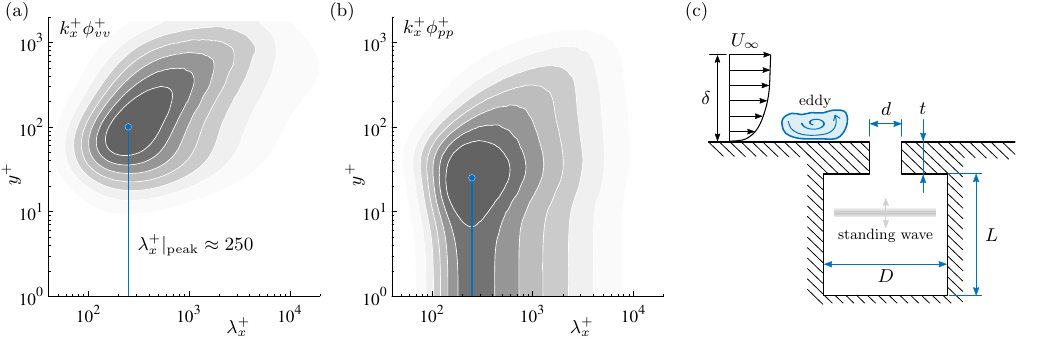}
\caption{Spectrograms of (a) the $v$ fluctuations, and (b) the $p$ fluctuations, as a function of the streamwise wavelength $\lambda_x$ and wall-normal distance $y$. These data are taken from a DNS of turbulent channel flow at $Re_\tau \approx 2\,000$ \citep{lee:2015a}. Grey-filled iso-contours correspond to levels of (a) $k_x^+\phi^+_{vv} = [0.05:0.05:0.35]$, and (b)  $k_x^+\phi^+_{pp} = [0.40:0.40:2.80]$. (c) Schematic illustration of a cylindrical HR beneath a grazing TBL flow.}
\label{fig:spectrointro}
\end{figure*}

Broadband chaotic motions in TBL flows are characterized by self-sustaining turbulence production mechanisms and transport phenomena. For any TBL, even at low $Re_\tau$ conditions, a self-sustaining cycle consists very near the wall \citep{kline:1967a,wallace:1972a,robinson:1991a,hwang:2013a} and its signature is well imprinted in energy spectra of the velocity fluctuations. In a premultiplied energy spectrogram of $u$, denoted as $k_x\phi_{uu}(\lambda_x,y)$, the well-known inner-spectral peak resides at $(\lambda_x^+,y^+) \approx (10^3,15)$, see Fig.~\ref{fig:TBLconditions}b. For the wall-normal velocity fluctuations, the peak energy in $k_x\phi_{vv}$ resides at a higher position and at smaller wavelengths, as seen from the peak at $(\lambda_x^+,y^+) \approx (250,100)$ in Fig.~\ref{fig:spectrointro}a. This spectrogram is inferred from Direct Numerical Simulation (DNS) data of turbulent channel flow \citep{lee:2015a} and corresponds to $Re_\tau \approx 2\,000$ (close to the Reynolds number of our current work). When considering the static pressure $p$, the primary hump of energy in $k_x\phi_{pp}$ resides at $(\lambda_x^+,y^+) \approx (250,25)$. Clearly, $k_x\phi_{pp}$ remains constant for $y^+ \lesssim 5$ and approaches the wall-pressure spectrum. Fluctuations in pressure and vertical velocity being most intense at a scale of $\lambda_x^+ \approx 250$ is relevant for our current work, since the pressure and velocity fields will excite and couple to the dynamics of the HRs to-be-studied.

In the near-wall region, energy in the spectrogram is tied to a re-occurring bursting process \citep{offen:1975a,blackwelder:1979a} including ejection (Q2) and sweep (Q4) events; the latter are responsible for the largest contribution to the turbulence skin friction \citep{orlandi:1994a,jimenez:1999a,fukagata:2002a}. With increasing $Re_\tau$, hairpin eddies become more pronounced, forming vortex packets \citep{adrian:2000a} and co-existing large-scale motions \citep{hutchins:2007ab}, all of which are interlinked \citep{marusic:2017a}. In the current work we confine ourselves to the near-wall turbulence. The wall-bounded turbulence community has a relatively firm understanding of how to control near-wall flow mechanisms responsible for generating skin-friction drag. For instance, much research has gone into the development of passive techniques, such as micro-textured riblet-surfaces and porous substrates \citep[][amongst others]{walsh:1983a,choi:1989a,bechert:2000a}. Riblets with an inner-scaled size of $\mathcal{O}(10l^*)$ reduce drag \citep{mayoral:2011a,krieger:2018a,modesti:2021a}, whereas larger riblets lead to an increase in drag due to roughness effects. Benefits similar to those provided by properly sized riblets can potentially be achieved with anisotropic porous substrates, comprising a wall-normal permeability much lower than the ones in the streamwise and spanwise directions \citep{abderrahaman:2017a,rosti:2018a,gomez:2019a}. Even though manufacturing these kind of surfaces in a way that those prevent spanwise rollers, resembling Kelvin–Helmholtz vortices, has not yet been proved practical \citep{chavarin:2020a}. Nevertheless, this concept still offers promising research pathways for advancing innovative surfaces.

Here we explore another near-wall flow control concept, revolving around a sub-surface unit-texture being an inner-scaled HR. A trail of recent studies attempted to attenuate Q4 sweep events using wall-embedded micro-cavities (circular blind holes) in wall-bounded turbulence $Re_\tau \approx 1660$ \citep{silvestri:2017a,silvestri:2017ab,silvestri:2018a}. \citet{choi:1993a} showed that a single rectangular cavity of size $180 < d^+ < 240$ (where $d^+$ is the dimension of the cavity cut-out) can yield a small net drag reduction. \citet{silvestri:2017ab} hypothesised based on work of micro-perforated panel absorbers \citep{maa:1998a} that cavities could potentially \emph{``dampen the coherent structures and disrupt the bursting cycle, which is responsible for the shear stress and viscous drag [...]"}. After examining the sweep events through variable interval time-averaging (VITA) analysis of $u$ time series, for a range of cavity diameters ($20 < d^+ < 150$), a case with $d^+ \approx 60$ resulted in an optimum attenuation of the streamwise TKE (reduction of roughly 13\,\%). It is important to state that most studies of micro-cavity arrays \citep{silvestri:2017a,silvestri:2017ab,silvestri:2018a,severino:2022a,scarano:2022a} solely rely on time series of $u$, generated using hot-wire anemometry at a few downstream locations of the array, and thus do not include spatial velocity fields and/or direct measures of the skin-friction drag. Hence, no firm conclusions regarding drag can be made (\emph{e.g.}, indirect methods such as assessing the shift in the logarithmic region of the mean velocity profile suffers from the development of an internal boundary layer). Likewise, VITA analyses based on the time series of $u$ are hard to interpret, since thresholds used in the analyses are based on the time series and these themselves change between the uncontrolled and controlled flows. In follow-up research to the micro-cavities with blind holes, \citet{silvestri:2017a} and \citet{bhat:2021a} found that a single backing cavity---to which all holes were connected---could further reduce the streamwise TKE when increasing the volume of the backing cavity. However, no explanation of the underlying mechanism was provided. 

\subsection{Fundamentals of Helmholtz resonators}\label{sec:introHR}
The current work approaches the study of sub-surface cavities from a fundamental angle by concentrating on a single HR. A cylindrical design is shown in Fig.~\ref{fig:spectrointro}c and its geometry is characterized by four parameters: the orifice diameter $d$ and thickness $t$ of the neck, and the diameter $D$ and depth $L$ of the cavity. Helmholtz formulated an expression for the resonance frequency when subject to acoustic pressure waves (in the absence of grazing flow):
\begin{equation}\label{eq:HR1}
    f_r = \frac{a_0}{2\pi}\sqrt{\frac{S}{V_c\left(t + t^*\right) + P}}.
\end{equation}
Here, $S = \pi d^2/4$ is the area of the orifice and $V_c = \pi D^2 L/4$ is the volume of the cavity. Parameter $a_0$ signifies the sound speed. Terms $P$ and $t^*$ are `end-corrections' and account for the fact that resonance displaces not only the fluid medium within the neck, but also a small portion of fluid inside the cavity and outside of the orifice. Different end-corrections can be found in the literature on acoustic resonators \citep[\emph{e.g.},][]{ingard:1953a,chanaud:1994a}. Based on a wave-tube analysis, \citet{panton:1975b} showed that Eq.~\eqref{eq:HR1} with $P = 0$ is only valid when $L$ is smaller than 1/16$^{\rm th}$ of the acoustic wavelength. To account for longer cavities, they proposed the correction term $P = L^2S/3$ and this one we adopt in the current work. The lumped end-correction of \citet{ingard:1953a}, $t^* = t^*_{\rm in} + t^*_{\rm out} = 0.48\sqrt{S}(1-1.25d/D) + 0.48\sqrt{S}$ (valid for $d/D < 0.4$), was shown to work well for design and is also adopted here (note that $t^*_{\rm in}$ and $t^*_{\rm out}$ refer to the portions of fluid inside the cavity and outside of the orifice, respectively). All in all, the true end-correction may vary based on the orifice geometry, cavity geometry and properties of the grazing flow.

\begin{figure*}[htb!] 
\vspace{0pt}
\centering
\includegraphics[width = 0.999\textwidth]{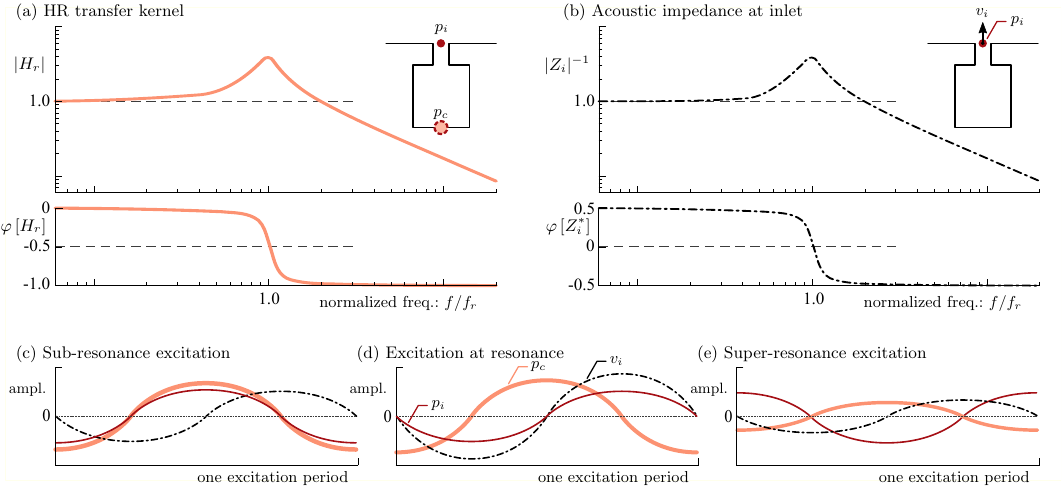}
\caption{Bode plot with gain and phase of (a) the HR transfer kernel between the inlet and cavity pressures, and (b) the acoustic impedance at the neck-inlet between the co-existing pressure and vertical velocity. (c,d,e) Illustrations over one excitation period, of the temporal fluctuations of $p_i$, $p_c$ and $v_i$ for a pure acoustic resonator with sub-resonance excitation, excitation at resonance, and super-resonance excitation, respectively.}
\label{fig:HRimped}
\end{figure*}

A HR results in amplitude and phase relations between the pressure at the neck-inlet and in the cavity ($p_i$ and $p_c$ in the inset of Fig.~\ref{fig:HRimped}a). These relations can be modelled through a single input/output transfer kernel, $H_r$, by considering a mass-spring-damper system analogy of a HR. Its gain and phase are given by,
\begin{equation}\label{eq:HRTF}
    \vert H_r\left(f\right)\vert = \left[\left(1-\left(\frac{f}{f_r}\right)^2\right)^2 + \left(\frac{2\xi f}{f_r}\right)^2\right]^{-\nicefrac{1}{2}},~\text{and}~\varphi\left[H_r\left(f\right) \right] = -\tan^{-1}\left[\frac{2\xi\left(f/f_r\right)}{1-\left(f/f_r\right)^2}\right].
\end{equation}
Here, $\xi$ is the damping constant and $f_r$ is the resonance frequency. While $\xi$ must be determined empirically, $f_r$ can be estimated using Eq.~\eqref{eq:HR1}. Though, after conducting a calibration experiment with an acoustic pressure-excitation (\emph{e.g.}, a broadband noise field) and simultaneous sampling of the inlet pressure $p_i$ (input) and cavity pressure $p_c$ (output), both $\xi$ and $f_r$ can be inferred from fitting Eq.~\eqref{eq:HRTF} to the gain and phase of the complex-valued kernel,
\begin{equation}\label{eq:HRkernel}
    H_r^{\rm aco} = \frac{\langle P_c\left(f\right)P^*_i\left(f\right)\rangle}{\langle P_i\left(f\right) P^*_i\left(f\right)\rangle}.
\end{equation}
Here the numerator is the input-output cross-spectrum and the denominator is the input spectrum. A capital symbol indicates the Fourier transform, \emph{e.g.}, $P_c(f) = \mathcal{F}\left[p_c(t)\right]$. A sample bode plot of Eq.~\eqref{eq:HRkernel} is drawn in Fig.~\ref{fig:HRimped}a. The amplitude response peaks around the resonance frequency. When the resonator is subject to a harmonic wave at a frequency $f \ll f_r$, the cavity pressure responds in-phase and the gain tends towards unity for low frequencies. An out-of-phase behavior is present for $f \gg f_r$ . With increasing frequency, the fluid medium is subject to more friction, particularly in the neck, causing the gain to decrease.

Finally, a characteristic acoustic impedance at the neck-inlet (this local wall-impedance is denoted as $Z_i$) is important to consider in terms of how the velocity disturbance of an acoustic resonator interacts with the pressure field. Here we use the definition $Z_i \equiv P_i(f)/V_i(f)$, where $V_i(f) = \mathcal{F}\left[v_i(t)\right]$ and in our convention $v_i > 0$ corresponds to an upward neck-inlet velocity (a motion of the mass of air in the HR-neck in the positive $y$ direction; this is opposite to the convention in the acoustic literature). A bode plot of $Z_i$ is drawn in Fig.~\ref{fig:HRimped}b. At sub-resonance excitation ($f \ll f_r$), the inlet pressure is compliant with the inlet velocity: when the mass of air moves inside ($v < 0$) the air is compressed and the applied inlet pressure is a quarter-period ahead of the vertical velocity. Conversely, at super-resonance excitation ($f \gg f_r$), acoustic inertance manifests itself and the applied inlet pressure is a quarter-period behind the vertical velocity. At resonance, $\vert Z_i \vert$ is low and maximum flow into and out of the resonator occurs with zero-phase delay between $p_i$ and $v_i$. Temporal fluctuations of $p_i$, $p_c$ and $v_i$ are summarized in Figs.~\ref{fig:HRimped}c-e for a pure acoustic resonator with sub-resonance excitation, excitation at resonance, and super-resonance excitation, respectively. Recall that the acoustic wavelengths are much longer than the size of the HR itself (also true for our resonators). Consequently, the cavity pressure has a similar phase anywhere inside the sub-surface cavity, although we indicate it as the pressure at the cavity bottom (this is where it is measured experimentally).

\subsection{Helmholtz resonators for flow control}\label{sec:controlHR}
Controlling the near-wall region of a grazing flow with a HR has been the subject of laminar flow studies \citep[][amongst others]{michelis:2023aHR,michelis:2023aPC}. These studies were aimed at attenuating Tollmien-Schlichting (TS) waves. Incoming TS waves are nearly harmonic in nature and, consequently, the interaction between their convective pressure and velocity fluctuations, and the response-perturbations of the control device, is relatively well-defined. \citet{michelis:2023aHR} concluded that the pressure perturbations in the HR behave acoustically and that $f_r$ is unaltered when excited by TS waves rather than acoustic waves. Furthermore, only an amplification of the TS waves was achieved with a single degree-of-freedom HR. This is because the wall-normal velocity perturbation of the flow-excited resonator and the wall-normal velocity of the incoming TS wave are, on average, not opposing one another. Phononic crystals offer a more promising route in this regard, since a wider design space allows for tuning the phase between free-face surface displacements and the unsteady pressure at the wall \citep{willey:2023a}. 

For TBL flow grazing a HR, a wide variety of studies exist in terms of HR size relative to the boundary layer scales. Wall-bounded flows are often studied over (resonating) cavities that span multiple $\delta$ \citep{nelson:1981a,inagaki:2002a,dequand:20032a,ma:2009a,ghanadi:2014a,ghanadi:2015a,stein:2019a}. Response of the boundary layer flow downstream is not the focus of such studies, but rather properties of the Kelvin-Helmholtz (KH) instability of the shear layer, of the self-sustained Rossiter modes \citep{rossiter:1964tr}, or of cavity pressure-modes. The aforementioned studies do not consider tuning the HR to characteristics of the incoming flow. \citet{panton:1975a} were the first to do so by tuning relatively small HRs to different portions of the wall-pressure spectrum while exclusively focusing on their acoustic response (more details in \S\,\ref{sec:sizing}). \citet{flynn:1990a1} and \citet{flynn:1990a2} did extend the work to infer how the HR affected the TBL flow. They observed increased Reynolds stresses $\overline{u^2}^+$ and $\overline{v^2}^+$, and smaller changes in the Reynolds shear stress $-\overline{uv}^+$. No further data analysis exists on the attenuation/amplification of specific ranges of scales, or a phenomenological description of the flow interacting with the HR. 

Only very few studies do consider distributed wall-impedance conditions; arrays of HRs for separation control are an example \citep{yang:2013a,bodart:2017c}. However, many works are outer-scaled and designed for (global) flow separation control \citep[\emph{e.g.},][]{seifert:1993a,greenblatt:2000a,urzynicok:2002c,seifert:2004a}. Frequency-tuned surfaces for passive flow control of wall-bounded turbulence are also studied with resolvent analyses, in combination with impedance wall-boundary conditions \citep{jafari:2023a} and properties of compliant surfaces \citep{luhar:2015a}. These theoretical surfaces show promising results in that they can both suppress the near-wall cycle and larger-scale motions, however, (meta)materials would have to be developed to obtain such surface properties, if at all possible.

\subsection{Present contribution and outline}\label{sec:outline}
In summary, the flow physics corresponding to a grazing TBL flow over an inner-scaled, single degree-of-freedom HR are not well-understood. We aim to address to what degree an inner-scaled HR can be excited by a grazing TBL flow, and how the flow couples and reacts to a passively-established change in the local wall-impedance. Such scientific insight could inspire a new category of control surfaces. Contrary to the TS wave interaction \citep{michelis:2023aHR}, the incoming pressure and velocity fluctuations are broadband and possess a non-deterministic phase. This will presumably result in a dynamic response of the resonator dominated by a resonance-excitation type of behavior (Fig.~\ref{fig:HRimped}d). To our best knowledge, no detailed study has addressed the phase response of properly-scaled resonators for near-wall flow control.

This work is structured as follows. First a brief overview of the experimental methodology is provided in \S\,\ref{sec:expmethod}, as well as the characteristics of our baseline TBL flow. Then in \S\,\ref{sec:sizing} we cover the design philosophy for our inner-scaled resonators, followed by an overview of their response to TBL flow- and acoustic-excitation scenarios in \S\,\ref{sec:HRresponse}. Details on the response of the TBL flow are presented in \S\,\ref{sec:TBLresponse}.

\section{Turbulent boundary layer flow and experimental methodology}\label{sec:expmethod}
\subsection{Facility and instrumentation}\label{sec:facility}
Experiments were carried out in an open-return wind tunnel facility (W-tunnel) at the Delft University of Technology, equipped with a test section for TBL studies. A brief overview of the facility and instrumentation used is here provided, while details can be found elsewhere \citep{baars:2023a}. 

\begin{figure*}[htb!] 
\vspace{0pt}
\centering
\includegraphics[width = 0.999\textwidth]{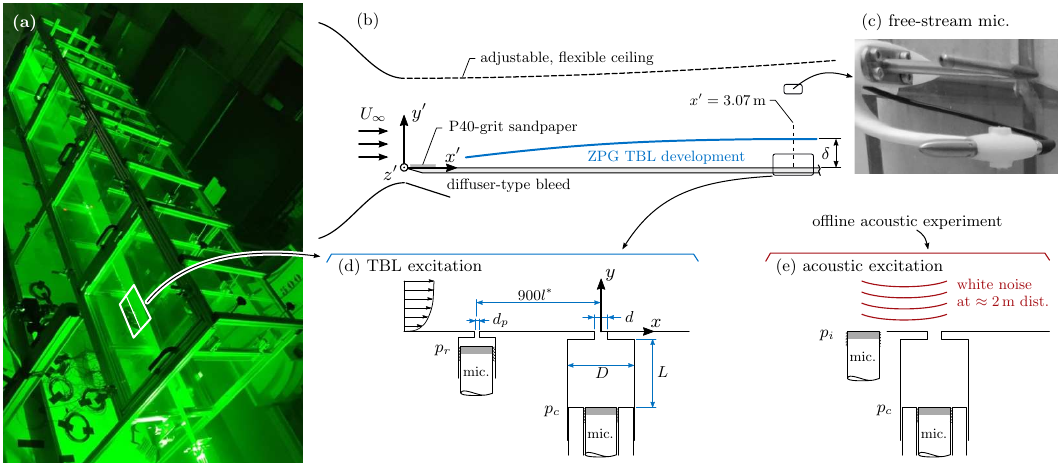}
\caption{(a,b) Setup for the TBL studies in the W-tunnel. (c) Photograph of the Pitot-static tube and microphone situated in the potential flow region. (d) Detail of a wall-embedded HR at $x = 0$, and a pinhole-mounted microphone located upstream for wall-pressure measurements. (e) Arrangement for acoustic characterization of a HR.}
\label{fig:TBLsetup}
\end{figure*}

For generating the TBL flow, a setup with a relatively long streamwise development length was employed (Figs.~\ref{fig:TBLsetup}a,b): the bottom wall over which the boundary layer developed under zero-pressure gradient (ZPG) conditions has a length of 3.75\,m and spans 0.60\,m in width, and a ZPG was achieved with the aid of an adjustable ceiling. The TBL was initiated just downstream of the leading edge at $x' = 0$, with a P40-grit sandpaper-trip over a length of 0.115\,m and over the full perimeter of the test section (all four surfaces). Measurements are performed near the downstream end of the setup, around $x' = 3.07$\,m. A coordinate system $(x,y,z)$ is used for the presentation of results in later sections, and has its origin at the midpoint of the HR orifice (see Fig.~\ref{fig:TBLsetup}d).

Single HRs were flush-mounted at the spanwise center of the test section, in a circular, modular plate with a diameter of 0.2\,m. This plate also embedded a pinhole-mounted microphone located $900l^*$ upstream, for measuring the incoming wall-pressure spectrum (details of which are provided in \S\,\ref{sec:HRresponse}). Given this spacing and the diameter of the pinhole of $d_p = 0.5$\,mm ($d_p^+ \approx 16$), the HR is exposed to an undisturbed inflow. Finally, a Pitot-static tube was mounted on the side-wall around $(x',y',z') = (3.07,0.30,-0.30)$\,m, for measuring the free-stream velocity $\overline{U}_\infty$; the static temperature and atmospheric pressure were also recorded for computing air density and viscosity. 

Fluctuating pressure measurements were conducted with three sensors: with a pressure-microphone at the bottom of the HR cavity to measure the cavity oscillations ($p_c$), with the pinhole-mounted microphone upstream to measure the wall-pressure spectrum ($p_r$) and in the potential flow region, near the Pitot-static tube (see photo in Fig.~\ref{fig:TBLsetup}c), to record facility noise ($p_f$). All three microphones were GRAS\,46BE $\nicefrac{1}{4}$\,in. CCP free-field microphones, with the grid caps removed. Only for the microphone measuring $p_f$, a GRAS\,RA002 nosecone was added to remove as much as possible any pressure fluctuations from free-stream turbulence. The microphone sets have a nominal sensitivity of 3.6\,mV/Pa and a frequency response range with an accuracy of $\pm 1$\,dB for $10$\,Hz to $40$\,kHz. The dynamic range is 35\,dB to 160\,dB (with a reference pressure of $p_{\rm ref} = 20 $\,$\upmu$Pa). Individual sensitivities were applied after inferring them with a GRAS\,42AG calibrator.

Boundary layer turbulence was captured using both hot-wire anemometry (HWA) and particle image velocimetry (PIV) measurements. For the former, a TSI IFA-300 bridge was used in CTA mode together with a Dantec 55P15 miniature-wire boundary layer probe (overheat ratio of around 1.8). Its sensing length is $l = 1.25$\,mm ($l^+ \approx 42.4$) and comprises a plated Tungsten wire with a diameter of $d_w = 5$\,$\upmu$m ($l/d_w = 250$). This wire provides a sufficient measurement resolution \citep{hutchins:2009a}, given that this study concentrates on control differences and not at providing fully-resolved turbulence measurements. Boundary layer profiles were acquired by traversing the hot-wire with a system possessing a position accuracy better than $0.4l^*$. One full profile was acquired with 40 points logarithmically spaced in the range $y^+ \in (10,2\,800)$, at $x^+ \approx -1\,695$, without HR to infer baseline TBL parameters (described in \S\,\ref{sec:TBLcharac}). When the HRs were embedded, profiles were acquired in the near-wall region only at $x^+ \approx 186$, with 20 points logarithmically spaced in the range $y^+ \in (7,100)$. For each hot-wire position, all time series of HWA voltage (and the microphone signals) were sampled at a rate of $\Delta T^+ \equiv \overline{U}_\tau^2/\nu/f_s = 0.36$, where $f_s = 51.2$\,kHz is the sampling frequency. Sampling was performed with a 24-bit A/D conversion and for an uninterrupted acquisition duration of $T_a = 60$\,seconds ($T_a\overline{U}_\infty/\delta \approx 13\,200$ boundary layer turnover times). Finally, the hot-wire probe was calibrated \emph{in-situ} using the reference velocity provided by the Pitot-static tube, and corrections for hot-wire voltage drift were implemented \citep{hultmark:2010a}.

Planar two-dimensional two-component (2D2C) PIV measurements were conducted in the $(x,y)$ plane. One LaVision Imager sCMOS camera was used with a 16-bit CCD sensor, with a size of $2\,560 \times 2\,160$\,px$^2$ and 6.5\,$\upmu$m pixel size. A Nikon lens with a 60\,mm focal length and an f\# of 8 was employed. A $1$\,mm thick sheet was illuminated by a double cavity Quantel Evergreen EVG00200 Nd:YAG laser, with a maximum energy per pulse of 200\,mJ. Flow tracers were generated with the aid of an atomized glycol-water mixture, yielding an average particle size of around $1$\,$\upmu$m. Measurements captured a field-of-view (FOV) surrounding the HR orifice at $x = 0$, with a streamwise and wall-normal extent of roughly $1.45\delta$ and $1.20\delta$, respectively (image resolution of $26$\,px/mm). Image pairs were acquired with a frame-delay of 35\,$\upmu$s, and for each dataset a total of two sets of 2\,000 statistically independent images were acquired at a rate of 15\,Hz. For conditional averaging the 4\,000 velocity fields to phases in the cavity pressure oscillations in post-processing, the laser Q-switch of the first pulse was acquired synchronously with the microphone signals $p_c$, $p_r$ and $p_f$ with the same DAQ specifications as mentioned above for the HWA acquisition. PIV processing was performed with LaVision DaVis 10.2 utilizing multi-pass cross-correlation. In the final pass, a 50\,\% overlap was employed with an interrogation window size of $12 \times 12$\,px$^2$, leading to a vector pitch of $0.23$\,mm ($7.8l^*$). 


\subsection{Turbulent boundary layer characteristics}\label{sec:TBLcharac}
\begin{figure*}[htb!] 
\vspace{0pt}
\centering
\includegraphics[width = 0.999\textwidth]{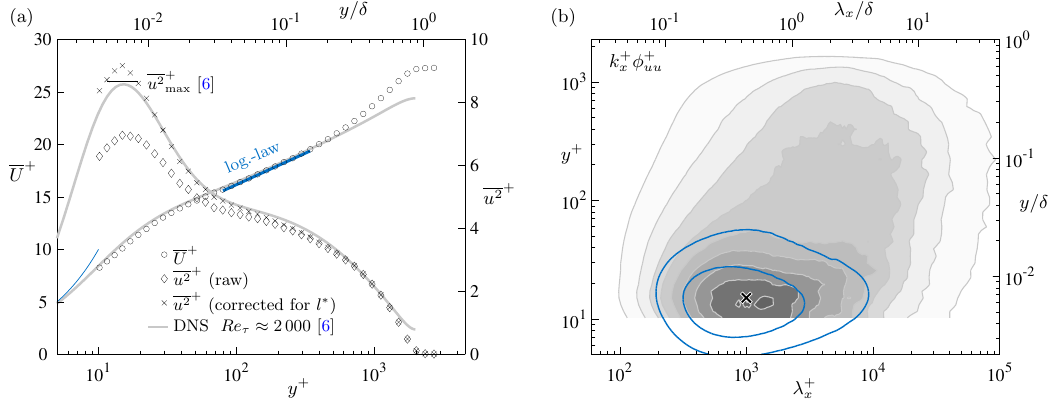}
\caption{Baseline properties of the TBL. (a) Streamwise MVP and streamwise TKE profile. Both are compared to DNS data of turbulent channel flow at $Re_\tau \approx 2\,000$ \citep{lee:2015a}. The MVP is compared to the logarithmic law with constants $\kappa = 0.384$ and $B = 4.17$. The streamwise TKE profile is corrected for spatial resolution effects. (b) Premultiplied energy spectrogram $k^+_x\phi^+_{uu}$ (filled iso-contours 0.2:0.2:1.8); note that $\lambda_x = \overline{U}_c(y)/f$. Overlaid are two iso-contours inferred from the DNS data and correspond to levels of $k_x^+\phi^+_{uu} = [1.0;~1.6].$}
\label{fig:TBLconditions}
\end{figure*}

Characteristics of the baseline TBL flow are now presented based on the HWA data. A mean velocity profile (MVP) and streamwise TKE profile are presented in Fig.~\ref{fig:TBLconditions}a. Boundary layer parameters are listed in Table~\ref{tab:expcon} and were obtained by fitting the MVP to a composite profile with log.-law constants of $\kappa = 0.384$ and $B = 4.17$ \citep{monkewitz:2023a}. On the basis of these values the friction Reynolds number is $Re_\tau \approx 2\,280$. Our MVP agrees well with the DNS data of turbulent channel flow at $Re_\tau \approx 2\,000$ \citep{lee:2015a}, up to the wake-region. For the streamwise TKE profile, an attenuation of the energy is observed due to the hot-wire's spatial resolution limit \citep{hutchins:2009a}. After correcting for the missing energy following the procedure of \citet{smits:2011a} the profile agrees well with the DNS profile in the buffer region and above. All in all, the data confirms a proper baseline flow at a practically relevant value of $Re_\tau$.

\begin{table}[htb!] 
\centering
\begin{tabular}{cccccccccc}
~$Re_\tau$~ & ~$Re_\theta$~ & ~$\overline{U}_\infty$ (m/s)~ & ~$\delta$ (mm)~ & ~$\theta$ (mm)~ & ~$\overline{U}_\tau$ (m/s)~ & ~$l^* \equiv \nu/\overline{U}_\tau$ ($\upmu$m)~ & ~$\nu/\overline{U}_\tau^2$ ($\upmu$s)~ & ~$\Pi$~ & ~$x^+$ location~\\ \hline \hline
2\,280 & 6\,190 & 14.8 & 67.3 & 6.71 & 0.54 & 29.50 & 54.56 & 0.56 & $-1\,695$\\ \hline \hline
\end{tabular}
\caption{Experimental parameters of the baseline TBL flow in the W-tunnel, inferred from a mean velocity profile acquired with HWA at $x = -50$\,mm (just upstream of where the HR is placed during control-experiments).}
\label{tab:expcon}
\end{table}

For evaluating the spectral content of the HWA data, one-sided spectra are taken as $\phi_{uu}(y;f) = 2\langle U(y;f) U^*(y;f)\rangle$, where $U(y;f) = \mathcal{F}\left[u(y,t)\right]$ is the temporal FFT. Here, the angular brackets $\langle \cdot \rangle$ denote ensemble averaging, and superscript $*$ signifies the complex conjugate. Ensemble averaging is performed using FFT partitions of $N = 2^{12}$ samples (subject to a Hanning window), resulting in a spectral resolution of ${\rm d}f = 12.5$\,Hz with 1\,500 ensembles and 50\,\% overlap. For interpretative purposes, frequency spectra are converted to wavenumber spectra using a convection velocity $\overline{U}_c(y)$, taken according to the local mean velocity except for $y^+ \lesssim 10$ where the convection velocity plateaus to a constant value of $\overline{U}^+_c = 10$ according to Fig.~3 of \citet{liu:2020a}. With streamwise wavenumber $k_x = 2\pi f/\overline{U}_c$ the premultiplied spectra, $k_x\phi_{uu}(k_x)$, are presented in terms of a wavelength on the scale axis, thus $\lambda_x = 2\pi/k_x = \overline{U}_c/f$. A streamwise energy spectrogram is shown in Fig.~\ref{fig:TBLconditions}b with a clear presence of the inner-spectral peak at $(\lambda_x^+,y^+) \approx (10^3,15)$.

\section{Resonator design and sizing}\label{sec:sizing}
When sizing a HR both spatial and temporal scales need to be decided upon. Here, we only consider a cylindrical HR (Fig.~\ref{fig:spectrointro}c). For the spatial scale the orifice diameter $d$ is the leading parameter because it dictates the size of the interface between the grazing flow and the sub-surface HR. A temporal scale is imposed by the resonance frequency, $f_r$. This work considers two inner-scaled HRs, meaning that $d$ and $f_r$ are chosen such that the HR interacts at resonance with the most energetic near-wall content \citep{jimenez:1999a}. Tuning of the resonator size and frequency to the near-wall scales ensures less drag in comparison to an outer-scaling. Note, however, that there is generally a drag increase in practical applications of acoustic liners, as reported in \citet{shahzad:2023a}, amongst others, though acoustic liners are scaled to the fan blade passing frequency of turbofan engines and are not explicitly scaled to the grazing TBL flow.

Regarding \textbf{\emph{spatial tuning}}, \citet{panton:1975a} emphasized that a maximum intensity of self-resonance is achieved when the turbulence excitation (\emph{e.g.}, a characteristic eddy size of a normal velocity fluctuation or pressure-event) is twice the orifice diameter, $d$. It was conjectured that this creates a pronounced \emph{in-flow} and \emph{out-flow} phase when convecting past the orifice, assuming that a shear-layer instability over the orifice does not play a major role. Here we proceed with the same assumption, but an objective of the current work is in fact to confirm whether HRs resonate according to acoustic theory. When excitation events are much smaller than the orifice diameter, multiple \emph{in-flow} and \emph{out-flow} phases would appear over the orifice and would average out to a weaker excitation, hence this scenario is not considered. Both $v$ and $p$ events are most energetic at a characteristic scale of $\lambda^+_x = 250$ (Figs.~\ref{fig:spectrointro}a,b), and in a view of perfect periodicity this equates to event sizes of $\Delta x^+ = 125$. Two HRs with orifice diameters of $d^+ = 68$ ($d = 2$\,mm) and $d^+ = 102$ ($d = 3$\,mm) are investigated, of which all parameters are listed in Table~\ref{tab:rescon}; we refer to these with \HR{1} and \HR{2}, respectively. For \HR{1}, the near-wall scales are almost twice the size of the orifice diameter following the case of \citet{panton:1975a}. The larger diameter orifice is still below the characteristic scale of an excitation event, and is of interest since a larger orifice diameter typically results in less viscous losses in the neck (for the same $t$) and thus a stronger resonance. Note that in preliminary studies we also investigated a HR with a smaller orifice diameter of $d^+ = 36$ ($d = 1$\,mm), but no pronounced response was observed in the TBL flow, hence this case is omitted. For manufacturing reasons, the neck thickness was kept constant at $t = 4$\,mm. 

\begin{table}[htb!] 
\centering
\begin{tabular}{ccccccccccc|cc|cc}
Resonator & ~$d^+$~ & ~$d$ (mm)~ & ~$t$ (mm)~ & ~$D$ (mm)~ & ~$L$ (mm)~ & ~Label~ & ~$f_r$ (Hz)~ & ~$f_r^+$~ & ~$\lambda_{x,r}^+$~ & ~$\lambda_a/L$~~~ & ~~~$f_r^{\rm aco}$~ & ~$\xi^{\rm aco}$~ & ~$f_r^{\rm tur}$~ & ~$\xi^{\rm tur}$~\\ \hline \hline
\multirow{3}{*}{\HR{1}} & \multirow{3}{*}{68} & \multirow{3}{*}{2.0} & \multirow{3}{*}{4.0} & \multirow{3}{*}{11.0} & 48.0 & \LF{} & 581 & 0.032 & 315 & 12.4 & 576 & 0.125 & 631 & 0.192 \\
     & & & & & 24.0 & \NF{} & 843 & 0.046 & 217 & 17.1 & 833 & 0.166 & 853 & 0.132 \\
     & & & & & 8.0 & \HF{} & 1\,485 & 0.081 & 123 & 29.1 & 1\,446 & 0.193 & 1\,385 & 0.145 \\ \hline
\multirow{3}{*}{\HR{2}} & \multirow{3}{*}{102} & \multirow{3}{*}{3.0} & \multirow{3}{*}{4.0} & \multirow{3}{*}{11.0} & 80.0 & \LF{} & 585 & 0.032 & 313 & 7.4 & 584 & 0.062 & 622 & 0.084 \\
     & & & & & 48.0 & \NF{} & 794 & 0.043 & 231 & 9.1 & 794 & 0.080 & 833 & 0.081 \\
     & & & & & 16.0 & \HF{} & 1\,454 & 0.079 & 126 & 14.9 & 1\,464 & 0.099 & 1\,413 & 0.084 \\ \hline \hline
\end{tabular}
\caption{Geometric parameters and resonance frequencies for each inner-scaled Helmholtz resonator (HR). Design frequencies $f_r$ are converted to a streamwise wavelength using $\lambda_{x,r}^+ \equiv \overline{U}^+_c/f^+_r$, where the convection velocity is taken as $\overline{U}^+_c = 10$. Cavity depths $L$ are compared to the acoustic wavelength $\lambda_a \equiv a_0/f_r$ in the column listing $\lambda_a/L$.}
\label{tab:rescon}
\end{table}

When considering \textbf{\emph{temporal tuning}}, the resonance frequency $f_r$ is easily adjusted by varying the cavity depth, $L$. For all resonators the cavity diameter is held constant at $D = 11$\,mm so that the cylindrical plug---closing the HR cavity from the bottom---allowed for embedding a pressure microphone for measuring the cavity pressure, $p_c$ (the microphone body has an outer-diameter of 6.35\,mm). Three different frequencies are considered. First, a nominal resonance frequency matches the dominant scale in $v$ and $p$ fluctuations ($\lambda_x^+ = 250$). With $\overline{U}_c^+ = 10$, the target frequency becomes $f^+ = \overline{U}^+_c/\lambda_x^+ = 0.04$. With $d$, $t$ and $D$ being fixed, $L$ is designed according to Eq.~\eqref{eq:HR1} with the correction terms $t^*$ and $P$ described in \S\,\ref{sec:introHR}. This case is referred to as the nominal frequency (\NF{}) case and Table~\ref{tab:rescon} lists all relevant parameters. For ease of manufacturing $L$ was rounded and results in slight changes in the design resonance frequency, $f_r^+$. For each resonator, two additional design frequencies are considered and these are referred to as low- and high-frequency cases (\LF{} and \HF{}, respectively). For the former, $L$ is adjusted such that the spatial wavelength induced is a factor 1.5 larger than the most energetic wavelength. This results in $f^+_r \approx 0.03$. Likewise, for the \HF{} case, $L$ is adjusted so that smaller spatial wavelengths are targeted (higher frequencies, $f^+_r \approx 0.08$). Our parameter sweep over frequency aims to test how the energetic near-wall cycle responds, on average, to an interaction with HRs that interact in/out-of-phase with different portions of the wall-pressure spectrum.

\section{Response of the Helmholtz resonators}\label{sec:HRresponse}
Resonators are assessed in terms of their response to pure acoustic excitation, as well as to excitation with grazing TBL flow. The acoustic experiment was conducted in the A-tunnel facility of the Delft University of Technology, which includes a test room that is anechoic at frequencies beyond 200\,Hz \citep{merino:2020a}. White noise was produced with a Bose\textsuperscript{\textregistered} speaker at a distance of approximately 2\,m from the HR, and was configured so that the acoustic wavefronts were co-planar with the orifice plane (Fig.~\ref{fig:TBLsetup}e). Noise recordings were made using two microphones similar to the ones described in \S\,\ref{sec:introHR}: one captured the cavity pressure, $p_c$, and one was flush-mounted next to the HR orifice to capture the neck-inlet pressure, $p_i$. With recordings of $T_a \approx 120$\,seconds, and ensemble-averaging using FFT partitions of $N = 2^{13}$ samples (1\,500 ensembles with 50\,\% overlap), a spectral resolution of ${\rm d}f = 6.25$\,Hz was obtained for the HR transfer kernel $H_r^{\rm aco}$ given by Eq.~\eqref{eq:HRkernel}.

\begin{figure*}[htb!] 
\vspace{0pt}
\centering
\includegraphics[width = 0.999\textwidth]{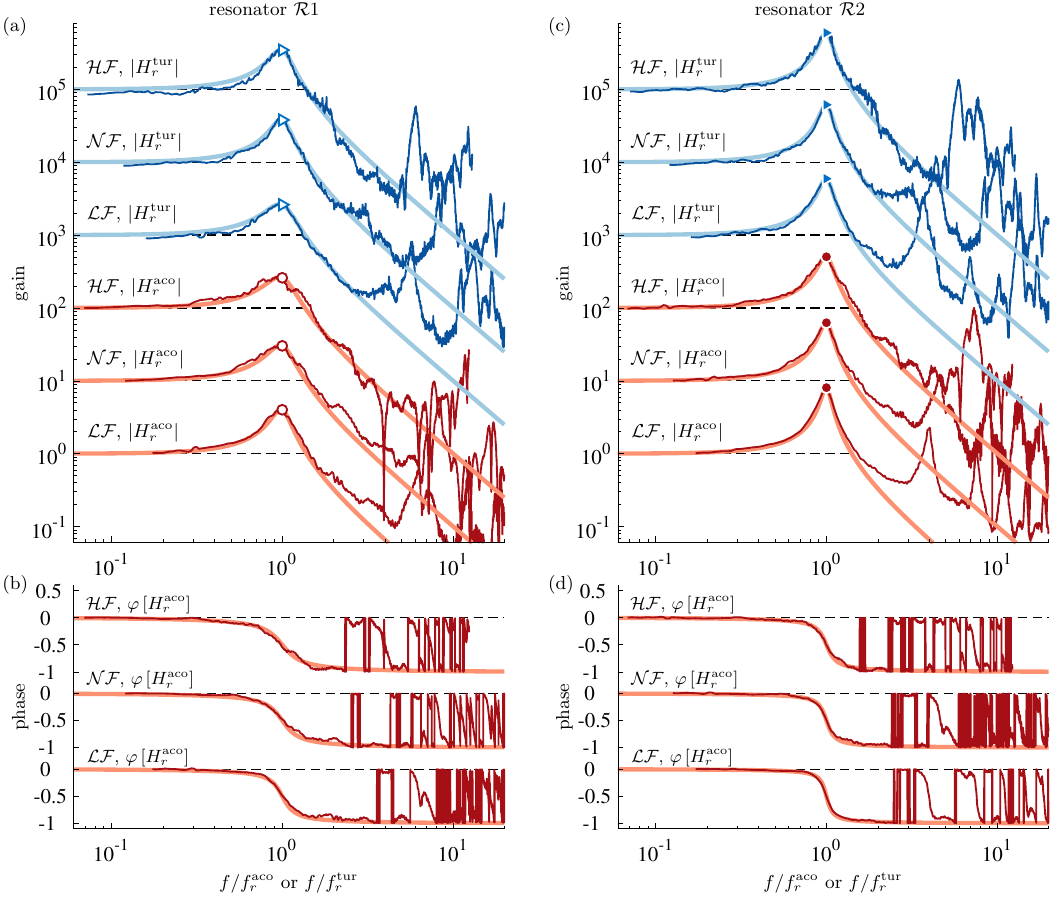}
\caption{(a,b) For resonator \HR{1}, bode plots of $H_r^{\rm aco}$ with the gain in sub-figure (a) and the phase in sub-figure (b), for the \LF{}, \NF{}, and \HF{} cases (series or red lines). In addition, the gain of $\vert H_r^{\rm tur} \vert$ is shown in sub-figure (a) with the series of blue lines. For all cases the experimental data is shown with a dark line, and a fit of the 2$^{\rm nd}$-order model transfer kernel to these data is shown with a thick light-shaded line. (c,d) Similar to sub-figures (a,b), but now for resonator \HR{2}.}
\label{fig:HRbode}
\end{figure*}

Bode plots of $H_r^{\rm aco}$ for resonator \HR{1} are provided in Figs.~\ref{fig:HRbode}a,b. The gain for all three resonance frequencies is presented with the three red curves at the bottom of Fig.~\ref{fig:HRbode}a, while the corresponding phase curves are shown in Fig.~\ref{fig:HRbode}b. Abscissae are normalized with the resonance frequency. Dark lines correspond to the measurements and are plotted for $f \gtrsim 100$\,Hz only due to the non-anechoic nature of the facility at lower frequencies. The gain of Eq.~\eqref{eq:HRTF} was fit to the measured gain as follows: first a peak frequency $f_p$ was identified, after which the gain expression in Eq.~\eqref{eq:HRTF} was fitted to the measured gain in the range $f \in [f_p/1.75,~1.75f_p]$. A nonlinear least squares fit was implemented with $f_r$ and $\xi$ as free parameters. After the fitting procedure the gain and phase of the model kernel were plotted with the light-shaded lines. Bode plots for resonator \HR{2} are presented in an identical manner in Figs.~\ref{fig:HRbode}c,d, and the inferred resonance frequencies and damping constants, $f_r^{\rm aco}$ and $\xi^{\rm aco}$, respectively, are listed in Table~\ref{tab:rescon}.

Overall, the model transfer kernel represents the data fairly well up to two-to-three times the resonance frequency, after which overtones are occasionally present (particularly in the \LF{} case of resonator \HR{2}). Analytical solutions of such overtones can be found by solving the acoustic wave-tube equations \citep{panton:1975b}. It was confirmed that the predicted frequencies agree well to the overtones in the measurements \citep{siebols:2022msc}. Nevertheless, since overtones have a much weaker gain and occur at frequencies beyond the energetic scales of the flow, these tones are not further considered.

\begin{figure*}[htb!] 
\vspace{0pt}
\centering
\includegraphics[width = 0.999\textwidth]{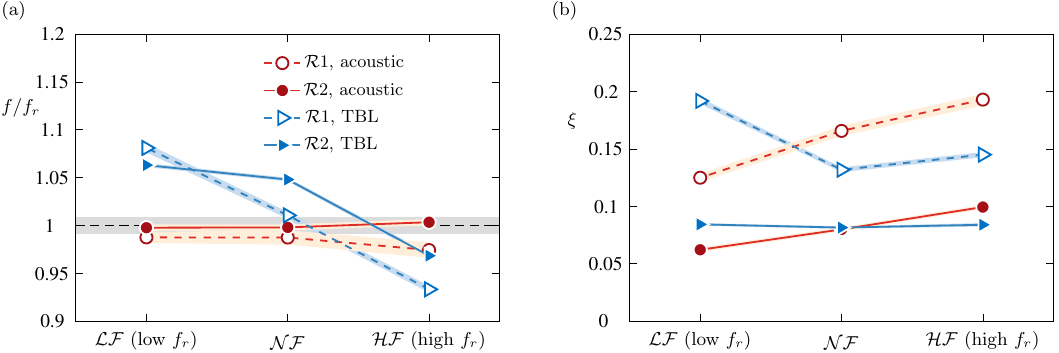}
\caption{(a) Empirically determined resonance frequencies in the case of acoustic excitation ($f_r^{\rm aco}$) and excitation by the grazing TBL flow ($f_r^{\rm tur}$), normalized with the design resonance frequency listed in Table~\ref{tab:rescon} ($f_r$), for the \LF{}, \NF{}, and \HF{} cases and for both resonators \HR{1} and \HR{2}. Similar to sub-figure (a), but now for the damping constant $\xi$.}
\label{fig:HRinferred}
\end{figure*}

Values of $f_r^{\rm aco}$ and $\xi^{\rm aco}$ are plotted in Figs.~\ref{fig:HRinferred}a,b for both resonators with the red curves. The frequency ordinate is normalized with the design frequency in Fig.~\ref{fig:HRinferred}a. A shaded band around $f/f_r = 1$ indicates the expected variation in the resonance frequency due to temperature changes of $\pm 5$\,K (this changes the sound speed and hence the resonance frequency). Shaded bands around the data curves correspond to the 95\,\% confidence interval of the nonlinear least squares to obtain $f_r^{\rm aco}$ and $\xi^{\rm aco}$. Clearly, the HRs under acoustic excitation behave according to the acoustic design formulation, Eq.~\eqref{eq:HR1}. Resonance frequencies deviate less than 3\,\%; such small discrepancies are ascribed to placement uncertainty of the cylindrical plug to adjust $L$. Damping constants follow the expected trends, in the sense that higher resonance frequencies and smaller neck diameter result in larger damping constants due to the acoustic particle velocities being subject to larger frictional losses.

Excitation of the HR response with the grazing TBL flow is considered next. The gain of the HR transfer kernel, $H_r^{\rm tur}$ (where superscript `tur' now refers to the turbulence excitation), is determined from the measured HR cavity pressure, $p_c$, and the upstream wall-pressure determined with the pinhole-mounted microphone signal, $p_r$. First of all, the combination of a pinhole and sub-surface cavity of the pinhole-mounted microphone behaves like a HR itself. Its transfer kernel $H_p^{\rm aco}$ was determined similar to the acoustic excitation experiment described above, and resonance was identified at roughly 4\,kHz (this is beyond the energetic temporal scales of interest in the wall-pressure). From the flow-on data, the gain $\vert H_r^{\rm tur} \vert$ can now be determined according to
\begin{equation}\label{eq:HRtur}
    \vert H_r^{\rm tur} \vert = \left(\frac{\langle P_c\left(f\right)P^*_c\left(f\right)\rangle}{\langle P_r\left(f\right) P^*_r\left(f\right)\rangle / \vert H_p^{\rm aco} \vert^2 }\right)^{\nicefrac{1}{2}} = \vert H_p^{\rm aco}\vert\left(\frac{\phi_{p_cp_c}\left(f\right)}{\phi_{p_rp_r}\left(f\right)}\right)^{\nicefrac{1}{2}}.
\end{equation}
The numerator equals the pressure spectrum of the HR cavity-pressure, while the denominator comprises the pressure spectrum of the pinhole-mounted microphone signal, corrected with the pinhole-cavity gain $\vert H_p^{\rm aco} \vert$ so that the numerator represents the pressure spectrum at the wall ($y = 0$). Note that a phase of $H_r^{\rm tur}$ cannot be determined, since this requires the neck-inlet pressure at $x = 0$ while the reference measurement cannot be co-located with the HR orifice. Note that Eq.~\eqref{eq:HRtur} assumes that the pinhole-cavity transfer kernel is invariant to the grazing flow (\emph{e.g.}, it was found from an acoustic calibration, although it is now embedded in a grazing flow) which is reasonable, given the small pinhole diameter ($d^+ \approx 16$); moreover, this is accepted as the conventional method for measuring TBL wall-pressure spectra \citep{willmarth:1975a}. As a final note, the facility noise, $p_f$, could be used to denoise the measurements of $p_c$ and $p_r$ \citep{baars:2023a}. However, in the division of Eq.~\eqref{eq:HRtur} this denoising procedure cancels and it was confirmed that the gain is unaltered.

Curves of the gain $\vert H_r^{\rm tur} \vert$ are plotted for both resonators in Figs.~\ref{fig:HRbode}a,c with the blue curves. After a fitting procedure identical to the one in the acoustic characterization, the resonance frequency, $f_r^{\rm tur}$, and damping constant, $\xi^{\rm tur}$, are plotted in Fig.~\ref{fig:HRinferred}a,b and listed in Table~\ref{tab:rescon}. It is apparent that the resonance frequency deviates more from the design frequency than in the case of acoustic excitation. For the \LF{} case, the resonance frequency is higher than the design frequency and this switches to the resonance frequency being lower than the design frequency for the \HF{} case (Fig.~\ref{fig:HRinferred}a). This suggests that end-correction terms $t^*$ and $P$ are altered by the grazing flow. Nevertheless, the deviation is less than 10\,\% for all cases, proving how Eq.~\eqref{eq:HR1} effectively valid for the design of a HR not only in quiescent conditions, but also for one subject to excitation with a grazing TBL flow (within an uncertainty of 10\,\%).

\section{Flow response of the turbulent boundary layer}\label{sec:TBLresponse}
\subsection{Statistical energy response of the turbulent boundary layer}\label{sec:TKEresponse}
\begin{figure*}[htb!] 
\vspace{0pt}
\centering
\includegraphics[width = 0.999\textwidth]{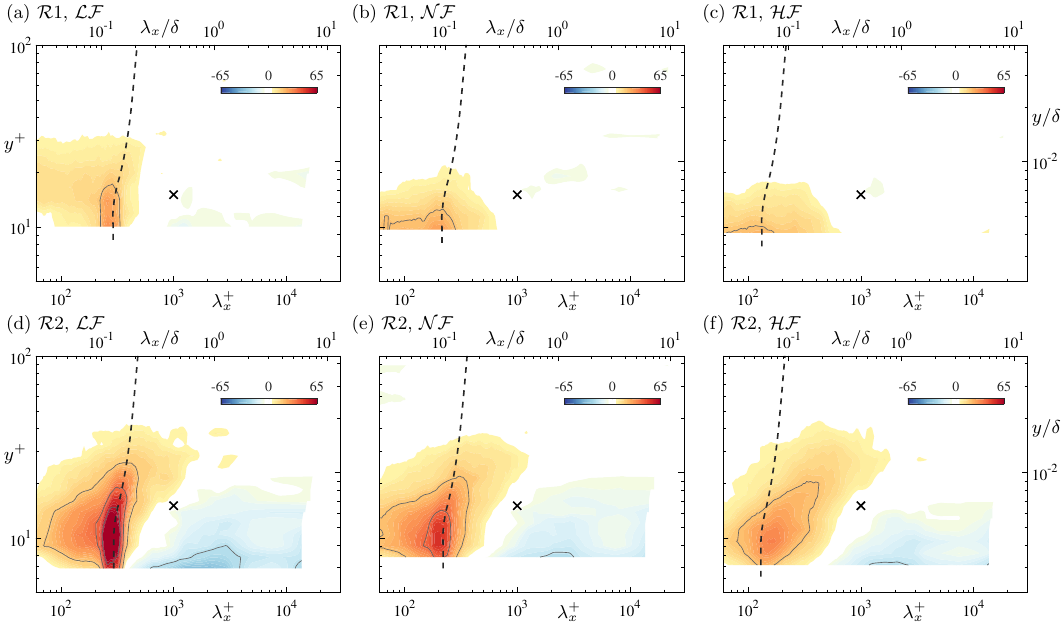}
\caption{Percentage difference of the premultiplied energy spectrograms $k_x^+\phi^+_{uu}$, for resonators \HR{1} (a-c) and \HR{2} (d-f) for all three resonance cases, at $x^+ \approx 186$, relative to the baseline spectrogram of the unperturbed flow presented in Fig.~\ref{fig:TBLconditions}b. Negative values (blue) indicate an energy reduction due to the presence of the HR, while positive values (red) indicate an energy intensification. Dark grey solid iso-contours correspond to levels of $\pm 20$\,\%, $\pm 40$\,\%, and $\pm 60$\,\%. A black dashed line corresponds to the constant resonance frequency; note that $\lambda_x = \overline{U}_c(y)/f$.}
\label{fig:GuuHWA23}
\end{figure*}

So far only the response of the HRs were analysed in terms of their transfer kernels between the neck-inlet and cavity pressures. How the HR affects the grazing TBL is assessed here, by first considering the change in the streamwise TKE captured accurately using HWA. For resonator \HR{1}, changes in the premultiplied energy spectrogram at $x^+ \approx 186$ are shown in Figs.~\ref{fig:GuuHWA23}a-c, for all three resonance cases. Changes are presented in terms of a percentage difference, relative to the baseline energy spectrogram presented in Fig.~\ref{fig:TBLconditions}b. Since HWA data provides frequency spectra, the difference was computed in the frequency domain after which all cases are presented with a unique conversion $\lambda_x \equiv \overline{U}_c(y)/f$, where $\overline{U}_c(y)$ is similar to the profile used in generating Fig.~\ref{fig:TBLconditions}b. A black dashed line corresponds to the constant resonance frequency and is located at $\lambda^+_{x,r}$ near the wall.

When inspecting the change in streamwise TKE, an increase is observed surrounding the resonance scale, which appears to be strongest in the \LF{} case. An increase of more than 20\,\% is observed in the buffer region (Fig.~\ref{fig:GuuHWA23}a), but the effect is only confined to $y^+ \lesssim 30$. For higher resonance frequencies (smaller scales), the increase in energy weakens as it becomes less narrow-band around the resonance scale. Changes in the streamwise TKE induced by resonator \HR{2} are larger compared to resonator \HR{1}, as seen in Figs.~\ref{fig:GuuHWA23}d-f. General observations are similar in that the largest increase of energy appears at the resonance scale for the \LF{} case (up to 70\,\% more energy) and this weakens when moving from the \LF{} to \HF{} case. In \S\,\ref{sec:CAresponse} we hypothesize a mechanism for this weakening effect. A stronger effect on the flow induced by resonator \HR{2}, in comparison to resonator \HR{1}, is consistent with the lower damping constants (Fig.~\ref{fig:HRinferred}b). Still, the energy change is confined to the buffer region $y^+ \lesssim 30$ and no disturbances penetrate the logarithmic region.

A major difference between resonators \HR{1} and \HR{2} is that for the \HR{2} resonators a strong attenuation of low-frequency energy is observed. More than 20\,\% of attenuation is present in the case of \LF{}, at scales on the order of the boundary layer thickness, \emph{e.g.}, $\lambda_x^+ = 2\,280$ (Fig.~\ref{fig:GuuHWA23}d). The mechanism associated with this decrease will be explored in \S\,\ref{sec:CAresponse}.

\subsection{Conditionally-averaged response of the turbulent boundary layer}\label{sec:CAresponse}
Given that \HR{2} caused a larger change in the streamwise TKE (in comparison to \HR{1}), we here proceed with the results of \HR{2} only. With the aid of the synchronized data of the cavity pressure, $p_c$, and the $u$ velocity from HWA, a conditional average of the velocity fluctuations is generated. 

\begin{figure*}[htb!] 
\vspace{0pt}
\centering
\includegraphics[width = 0.999\textwidth]{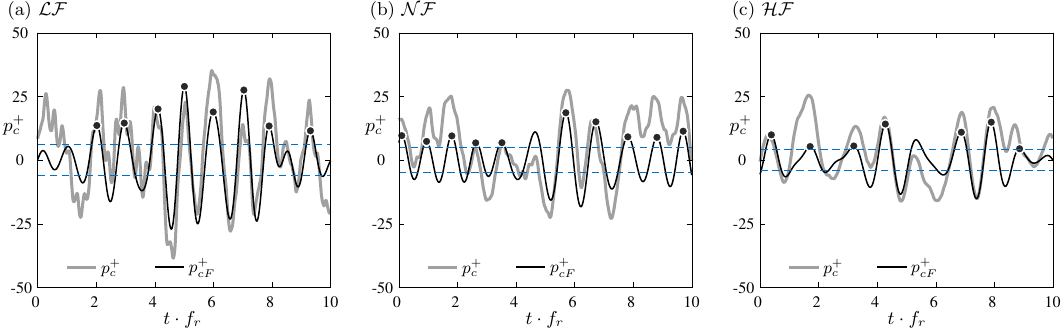}
\caption{Time series of the inner-normalized cavity pressure corresponding to \HR{2}, for the (a) \LF{} case, (b) \NF{} case, and (c) \HF{} case. Raw measurements are shown with the grey thick line, while a filtered version with a bandpass of $f \in [0.75f_r^{\rm tur},~1.50f_r^{\rm tur}]$ is shown with the black line. Blue dashed lines indicate $\pm 0.5p^+_{cF,{\rm rms}}$.}
\label{fig:pctime}
\end{figure*}

First, pressure time series of $p_c$ are shown in Fig.~\ref{fig:pctime} for all three resonance frequencies, and for a time span of 10 resonance periods. Raw time series are shown with the grey thick lines. Signals are not perfectly harmonic with resonance period $T_r = 1/f_r$, given that the wall-pressure excitation is broadband. Still, resonance dominates since $f_r$ roughly coincides with the dominant energetic frequencies of the wall-pressure spectrum. Moreover, the root-mean-square (rms) of the wall-pressure in this TBL flow is on the order of $p^+_{w,{\rm rms}} \approx 3.31$ following the empirical trend of \citet{klewicki:2008a}, to which our current TBL flow agrees \citep{baars:2023a}. The cavity pressure-rms is thus much higher than the wall-pressure-excitation-rms, \emph{e.g.}, $p_{c,{\rm rms}}/p_{w,\rm rms} \approx 5.42$ for the case of \LF{} (which relates close to the gain factor at resonance of $0.5/\xi^{\rm tur} \approx 5.9$).

\begin{figure*}[htb!] 
\vspace{0pt}
\centering
\includegraphics[width = 0.999\textwidth]{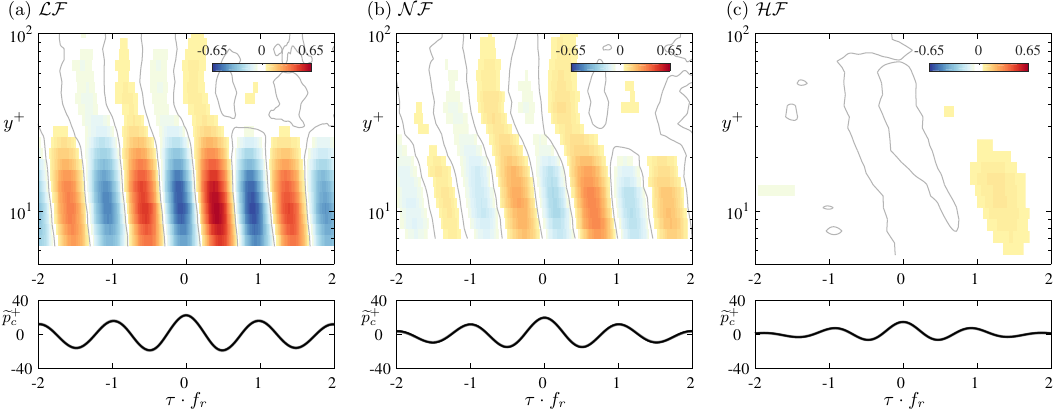}
\caption{Iso-contours of the conditionally-averaged streamwise velocity fluctuations, $\widetilde{u}^+(y,\tau)$, inferred from the HWA data at $x^+ \approx 186$, for the (a) \LF{} case, (b) \NF{} case, and (c) \HF{} case. Fluctuations are conditioned on the local maxima of the bandpass filtered cavity pressure signals (illustrated in Fig.~\ref{fig:pctime}); conditional averages of the (raw) cavity pressure signals themselves are shown in the bottom plots.}
\label{fig:CAHWA}
\end{figure*}

To perform conditional averaging accurately, raw time series are filtered with a narrow bandpass kernel around the resonance frequency, $f \in [0.75f_r^{\rm tur},~1.50f_r^{\rm tur}]$ to yield $p_{cF}$, shown with the black lines. Their local maxima are identified with a peak-detection algorithm, which only retains the maxima of which the amplitudes in both the filtered and raw time series exceed $0.5p^+_{cF,{\rm rms}}$ (shown with the blue dashed lines). Time series of $u$ are now conditionally-averaged to the local maxima, at each $y$ position, and are denoted as $\widetilde{u}(y,\tau)$ with $\tau = 0$ signifying the condition point. Fig.~\ref{fig:CAHWA} presents iso-contours of $\widetilde{u}(y,\tau)$ for a time span of 4 resonance periods, alongside the conditional average of the unfiltered cavity pressure, $\widetilde{p}_c$. As expected after inspecting the spectrograms in Fig.~\ref{fig:GuuHWA23}, the cavity pressure coherent $u$ fluctuations are strongest in the \LF{} case. When moving towards the \HF{} case, the magnitude decreases due to the lower level of pressure oscillations in the cavity. Moreover, the conditional average of the cavity pressure for the \HF{} case also shows a decaying amplitude away from $\tau = 0$, since the pressure is more broadband and the phase of resonance varies more erratically with time (recall Fig.~\ref{fig:pctime}c) than during the \NF{} or \LF{} case. This is also related to less gain in the \HF{} case (recall a higher $\xi$) and thus the magnitude of pressure resonance-oscillations is lower compared to the broadband content. Finally, the inclination of the velocity perturbations suggests that they are convective in nature and adapt to the mean shear in the TBL flow at the position of the measurement (a perturbation is first observed at positions further away from the wall, at $x^+ \approx 186$). In order to inspect the induced velocity perturbations and link them to the resonating pressure, we use spatial velocity fields from PIV data.

Conditional averaging of the 4\,000 PIV-based velocity fields---available for each resonator case---to the periodic resonance cycle, allows for a detailed inspection of the spatio-temporal evolution of the resonance perturbation. Local maxima in the cavity pressure (Fig.~\ref{fig:pctime}) occur at time instances denoted as $T_c$, and are identified with an identical procedure as the one used in conditional averaging the hot-wire time series. The resonance period $T_r$ is now divided into six equal-duration bins, each of which has a central instant $T_i$, with $i = 1 ... 6$. Given the synchronized acquisition of the laser Q-switch of the first pulse of the PIV image pair, with the cavity pressure data, each PIV-based velocity field is assigned to one of the 6 bins. It must be noted that bin 1 (and thus also the subsequent bins) is offset from the time instance of the local maxima in the cavity pressure signal, to account for two factors:\\[-14pt]
\begin{enumerate}[labelwidth=0.60cm,labelindent=0pt,leftmargin=1.00cm,label=(\roman*),align=left]
\item \noindent For resonators with relatively slender cavities of length $L$, a relatively small value of $\lambda_a/L$ (recall that $\lambda_a$ is the acoustic wavelength at resonance) causes a temporal lag between the measured pressure at the cavity bottom and the pressure in the near-vicinity of the orifice. As such, time instants $T_c$ are offset by $L/a_0$, forward in time, so that conditional averaging is done in relation to the pressure at the top of the cavity. This offset is relatively minor in the current work: the \LF{} case of \HR{2} has the deepest cavity with $\lambda_a/L \approx 7.4$ (Table~\ref{tab:rescon}) and this results in an offset of $L/a_0 = L/(a_0/f_r) T_r = 7.4 T_r \approx 0.14T_r$.\vspace{-6pt}
\item \noindent Given the $0.5\pi$ phase offset between the cavity pressure and the neck-inlet pressure at resonance, time instants of local maxima in the cavity pressure, $T_c$, are offset by $\nicefrac{1}{4}T_r$, also forward in time. This ensures that conditional averaging is done on a signal that is representative of the neck-inlet pressure (see Fig.~\ref{fig:HRimped}d). 
\end{enumerate}

\begin{figure*}[htb!] 
\vspace{0pt}
\centering
\includegraphics[width = 0.999\textwidth]{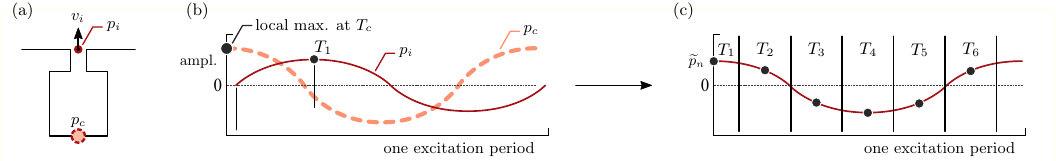}
\caption{Illustration of the bins used for conditional averaging of the PIV velocity fields. (a) Locations of the neck-inlet and cavity pressures, and (b) the phase-offset between the signals with the local maximum in $p_c$ at time instant $T_c$, and the central time instant $T_1$ of the first bin. (c) All six central time instants in relation to the signal representative of the neck-inlet pressure.}
\label{fig:PIVbins}
\end{figure*}

In short, the time instant of bin 1 corresponds to $T_1 = T_c + L/a_0 + 0.25T_r$ and signifies the maximum resonance pressure at the neck-inlet; this pressure signal is denoted as $\widetilde{p}_n$. Fig.~\ref{fig:PIVbins} illustrates the central time instants of all 6 bins in relation to the periodic signal $\widetilde{p}_n$. Recall that for determining the local maxima in the cavity pressure, a threshold was implemented to omit `weak resonance' periods. As such, for the \LF{} case as an example, 3\,112 fields of the 4\,000 fields in total were used for conditional averaging and were divided nearly equally over all six bins (484 fields for the bin with the lowest number of fields). Before averaging the velocity fields corresponding to each bin, a singular value decomposition (SVD) of the statistically-independent $u(x,y,t_i)$ and $v(x,y,t_i)$ snapshots was conducted in a combined $\left[u;v\right]$ fashion. Each field at $t_i$ was then regenerated through a low-dimensional reconstruction with a subset of modes (mode $m = 1...M$, where $M$ determined so that 40\,\% of the total resolved TKE was reconstructed). This step assisted in removing noise, but does not affect the conclusions made.

\begin{figure*}[htb!] 
\vspace{0pt}
\centering
\includegraphics[width = 0.999\textwidth]{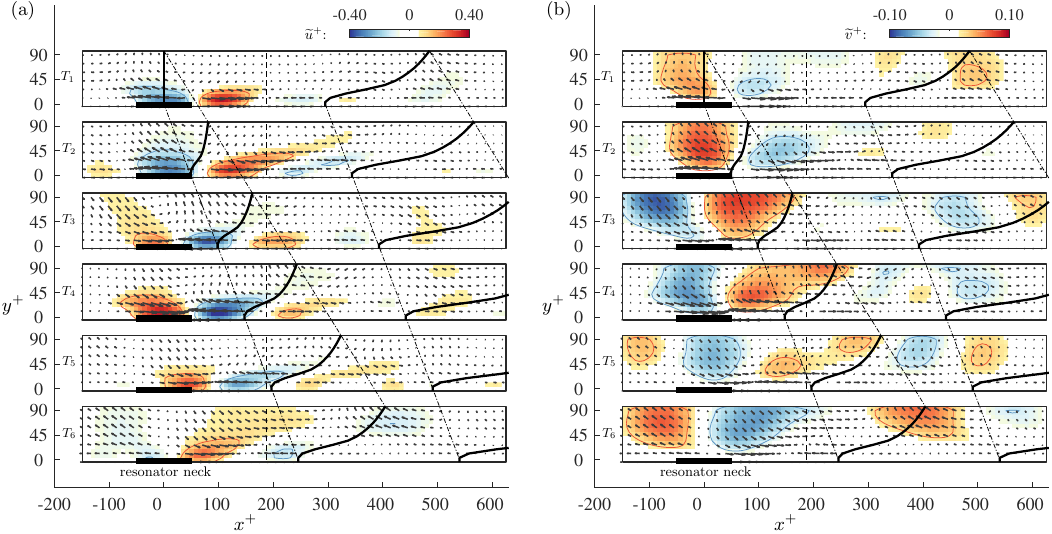}
\caption{Iso-contours of the conditionally-averaged (a) streamwise, $\widetilde{u}(x,y,T_i)$, and (b) wall-normal velocity fluctuations, $\widetilde{v}(x,y,T_i)$, for the \LF{} case of \HR{2}. Velocity fields are shown for all 6 instances, $T_i$ with $i = 1...6$, corresponding to the six central instants of each bin used for conditional averaging over one period, $T_r$ (see Fig.~\ref{fig:PIVbins}). The resonator neck of $d^+ = 102$ is indicated with the thick black line; a vertical line at $x^+ = 186$ indicates the hot-wire profile. A vertical line above the resonator in the $T_1$ field is displaced in each subsequent field according to the mean convection velocity profile, $\overline{U}_c(y)$, as described in the text. Because of periodicity in the conditional average, the visualization continues in frame $T_1$ after frame $T_6$.}
\label{fig:PIVLF}
\end{figure*}

Conditionally-averaged velocity fields for the \LF{} case are shown in Fig.~\ref{fig:PIVLF}a (contour of streamwise velocity $\widetilde{u}$) and Fig.~\ref{fig:PIVLF}b (contour of the wall-normal velocity $\widetilde{v}$). Fields are shown for each $T_i$, $i = 1 ... 6$, and vectors are superimposed on each field (with a vector skip of 2). Only a domain up to a wall-normal distance of $y^+ \approx 100$ is considered, given that our previous analysis already indicated a relatively limited wall-normal region influenced by the resonator. In streamwise direction, a total length of $\Delta x^+ \approx 750$ is considered and includes the resonator neck of which the extent is indicated with the thick line of length $d^+ = 102$. To accommodate the inspection of the convecting flow field during the resonance cycle, a vertical line above the resonator (in the field at $T_1$) is displaced in each subsequent field according to the mean convection velocity profile $\overline{U}_c(y)$, taken as the local mean velocity except for $y^+ \lesssim 10$ where the convection velocity plateaus to a constant value of $\overline{U}^+_c = 10$. Given the periodicity of the conditional average, this line continues to deform when moving from $T_6$ to $T_1$ and beyond.

Vertical velocity fluctuations $\widetilde{v}$ in Fig.~\ref{fig:PIVLF}b show an apparent outflow and inflow phase. These regions of positive and negative vertical velocity are accompanied by negative and positive streamwise velocity fluctuations $\widetilde{u}$, respectively (see Fig.~\ref{fig:PIVLF}a). This indicates that the the self-resonance disturbance is similar to one induced by zero-mass net flux jetting: during the inflow phase the high-momentum fluid in the TBL flow is pulled towards the wall, and vice versa during the outflow phase. It is intriguing to notice that the disturbance in the $\widetilde{u}$ velocity is primarily confined to the region $y^+ \lesssim 30$ (as was already concluded from the spectrograms in Fig.~\ref{fig:GuuHWA23}), while the disturbance in $\widetilde{v}$ reaches beyond $y^+ = 100$. This is evidence that the resonator is driven by, and locks in with, vertical velocity fluctuations that are themselves most energetic around $y^+ \approx 100$ as shown in the spectrogram of $v$ in Fig.~\ref{fig:spectrointro}a.

\begin{figure*}[htb!] 
\vspace{0pt}
\centering
\includegraphics[width = 0.999\textwidth]{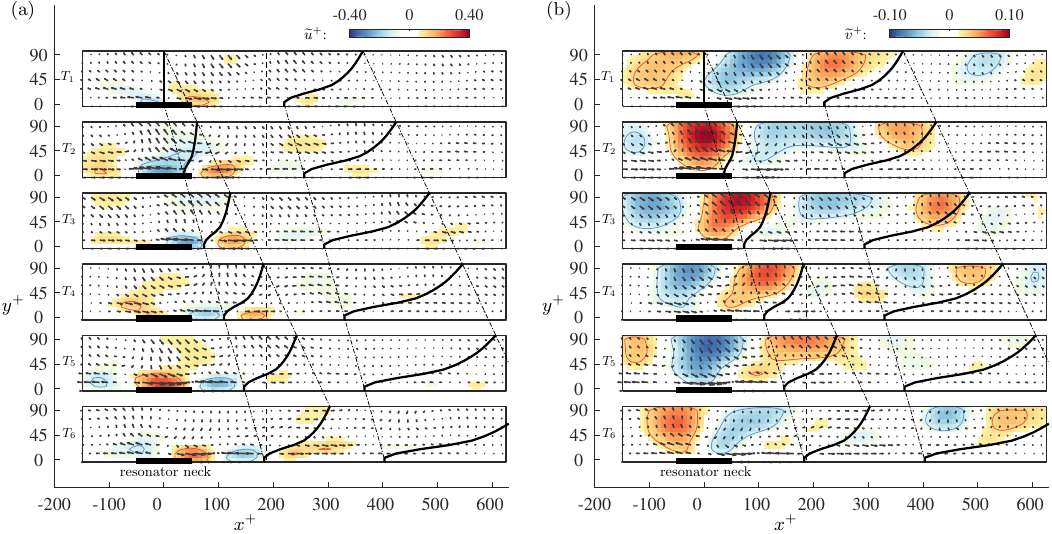}
\caption{Similar to Fig.~\ref{fig:PIVLF}, but now for the \NF{} case.}
\label{fig:PIVNF}
\end{figure*}

\begin{figure*}[htb!] 
\vspace{0pt}
\centering
\includegraphics[width = 0.999\textwidth]{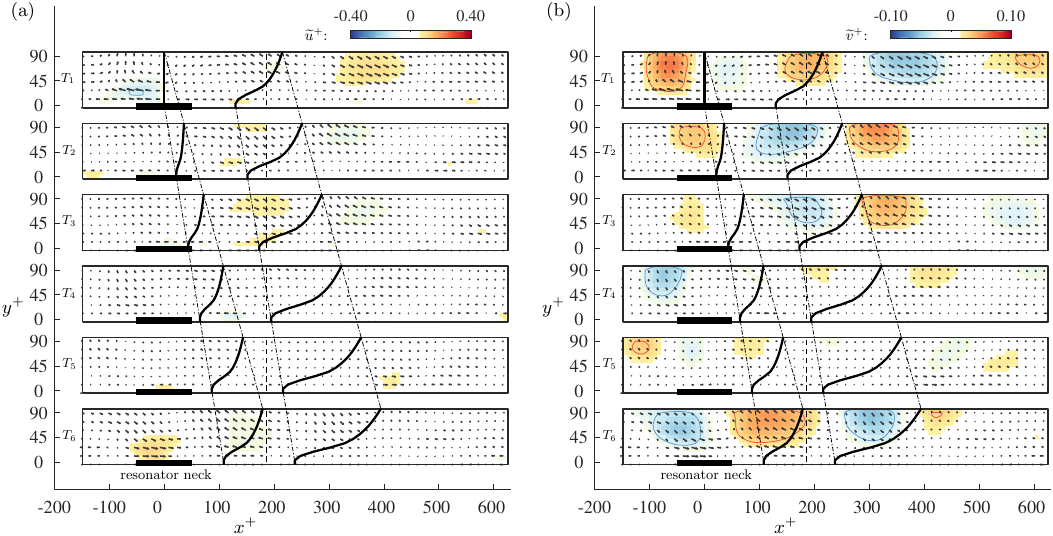}
\caption{Similar to Fig.~\ref{fig:PIVLF}, but now for the \HF{} case.}
\label{fig:PIVHF}
\end{figure*}

Before proceeding with flow-field observations, the conditionally-averaged fields for the two remaining resonance cases (\NF{} and \HF{}) are presented in Figs.~\ref{fig:PIVNF} and~\ref{fig:PIVHF}, and are shown in an identical format as the one of the \LF{} case in Fig.~\ref{fig:PIVLF}. For the \LF{} case, the velocity disturbance of the resonator persists downstream for a duration of roughly one resonance cycle. This is generally true for the \NF{} case also, although the $\widetilde{u}$ motions are slightly weakened while the $\widetilde{v}$ ones contain a more coherent character and persist downstream for almost two resonance cycles. This is ascribed to the fact that for the \NF{} case the resonator cycle is tuned to the period of the naturally-occurring $v$ motions that contain most energy and thus amplifies the most energetic fluctuations in the most distinguishable perturbation. To understand the significant weakening of the $\widetilde{u}$ (and $\widetilde{v}$) motions in the \HF{} case, it is important to consider the convective nature of the perturbations. We here define a stroke ratio with two time scales: (1) the half-period of resonance, $t_r \equiv T_r/2$, that is characteristic of the outflow and inflow duration, and (2) the convective time of the flow past the resonator's neck, $t_c \equiv d/\overline{U}_c$, with $\overline{U}_c^+ = 10$, yielding
\begin{equation}\label{eq:sr}
  {\rm SR} \equiv \frac{t_r}{t_c} = \frac{\overline{U}_c T_r}{2d}.
\end{equation}
For resonator \HR{2}, ${\rm SR} \approx 1.45$ for the \LF{} case, ${\rm SR} \approx 1.08$ for the \NF{} case and ${\rm SR} \approx 0.64$ for the \HF{} case. When ${\rm SR} > 1$, outflow and inflow phases are fully established throughout the entire neck orifice, \emph{e.g.}, an inflow signature associated with increased streamwise velocity convects over the entire orifice before it reverses to the outflow phase. When ${\rm SR} < 1$ (in the \HF{} case), an outflow and/or inflow phase does not convect past the orifice before the neck-inlet velocity changes sign. This reduces the effective diameter of the neck, \emph{e.g.}, near the trailing-edge of the neck the signatures of the outflow and inflow phases are weakened since they are continuously opposing one another.

\subsection{Pressure-velocity coupling at neck-inlet}\label{sec:impedance}
\begin{figure*}[htb!] 
\vspace{0pt}
\centering
\includegraphics[width = 0.999\textwidth]{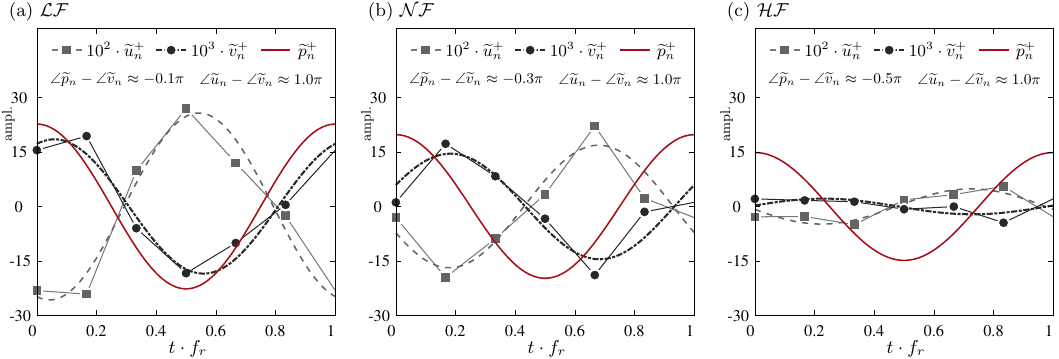}
\caption{Conditionally-averaged streamwise, wall-normal and pressure fluctuations over the orifice of \HR{2}, for the (a) \LF{} case, (b) \NF{} case, and (c) \HF{} case. For each of the 6 instances in the resonance cycle (corresponding to the fields shown in Figs.~\ref{fig:PIVLF}, \ref{fig:PIVNF} and \ref{fig:PIVHF}) the velocity fluctuations are averaged over a thin region above the orifice ($y^+ \approx 7$ and $x \in \left[-0.4d,0.4d\right]$) and shown with the markers; harmonic fits to the data are shown with the dashed and dash-dotted lines. A harmonic pressure oscillation reflects the pressure at the neck-inlet, inferred from the conditionally-averaged cavity pressure with the proper phase shifts applied (described in the text).}
\label{fig:necktime}
\end{figure*}

When time-resolved pressure information is available at the neck-inlet, together with time-resolved velocity fields, the local wall-impedance at the neck-inlet, $Z_i \in \mathcal{C}$ (as discussed in \S\,\ref{sec:introHR}), could be inferred from the data. However, given the absence of time-resolved data, an attempt is made to infer the impedance relation at the resonance scale only ($f = f_r$) from the conditionally-averaged behavior. The average streamwise and vertical velocities are extracted just above the neck-inlet at $y^+ \approx 7$, from the conditionally-averaged velocity fields (thus from the fields presented in Figs.~\ref{fig:PIVLF}a,b). These velocities are denoted as $\widetilde{u}_n$ and $\widetilde{v}_n$ and are plotted in Fig.~\ref{fig:necktime}a for each of the 6 instances in the resonance cycle of the \LF{} case; harmonic waves are fitted to the discrete data points and are described in the caption. In addition, a time series of the neck-inlet pressure is superimposed following the discussion of Fig.~\ref{fig:PIVbins}, in which it was pointed out that the velocities are conditionally-averaged so that the first bin of the resonance cycle corresponds to the maximum pressure at the neck-inlet. Figs.~\ref{fig:necktime}b and~\ref{fig:necktime}c are similar to Fig.~\ref{fig:necktime}a, but correspond to the \NF{} and \HF{} cases, respectively. 

Only the phase relation between $\widetilde{u}_n$, $\widetilde{v}_n$ and $\widetilde{p}_n$ can be interpreted, while their amplitude ratios (\emph{e.g.}, the ratio of the maximum in $\widetilde{p}_n$ to the maximum in $\widetilde{v}_n$ for determining $\vert Z(f \equiv f_r)\vert$) are non-physical given that velocities and pressure are based on conditional averaging, and the pressure magnitude is representative of the cavity pressure. Nevertheless, phase relations reveal that:\\[-14pt]
\begin{enumerate}[labelwidth=0.60cm,labelindent=0pt,leftmargin=1.00cm,label=(\roman*),align=left]
\item \noindent Vertical and horizontal velocities appear in perfect phase opposition: the phase shifts between the harmonic functions fitted to the data are $1.0\pi$ for all three resonance cases. This is direct evidence that at resonance, the near-wall horizontal velocity in the sheared TBL flow reacts instantly to the vertical perturbations displacing pockets of air with high-/low-speed streamwise velocity.\vspace{-6pt}
\item \noindent For the \LF{} case, the vertical velocity $\widetilde{v}_n$ and neck-inlet pressure are almost in-phase with only a $0.1\pi$ phase shift. For an acoustic resonator at resonance, this in-phase behavior is expected (recall Figs.~\ref{fig:HRimped}b and~\ref{fig:HRimped}d; $\varphi\left[Z_i^*\right] = 0$ at $f = f_r$). When moving towards the other resonance cases, this phase drifts towards a half-out-of-phase behavior for the \HF{} case: pressure is leading the vertical velocity, which would reflect sub-resonance behavior (Figs.~\ref{fig:HRimped}b and~\ref{fig:HRimped}e). We conjecture that this phase relation is due to the response delay in the vertical velocity: a significant part of the vertical perturbation near the trailing edge of the neck is attenuated.
\end{enumerate}

Finally, the attenuation of large-scale energy that was most noticeable in the \LF{} case (Fig.~\ref{fig:GuuHWA23}d) appears to be related to a mechanism at sub-resonance scale. For the \LF{} case it was confirmed that the HR behaves according to classical HR working mechanism reported in Fig.~\ref{fig:HRimped}. Presumably, at sub-resonance excitation, the resonator then induces a neck-inlet pressure that is leading the vertical velocity fluctuations. If the large-scale perturbations of the grazing TBL flow locks in with those of the resonator, then an opposition-type of mechanism is established: a larger percentage of the cyclic resonance opposes the motion than that it enhances the motion. However, time-resolved spatial velocity measurements, synchronized with neck-pressure information, must be conducted to confirm this claim and to expose the exact mechanism by which near-wall large-scale motions are attenuated.

\section{Concluding remarks}\label{sec:concl}
This work considered single, miniature HRs beneath grazing TBL flow. Response details of the resonator itself, as well as the TBL receptivity to the resonance, were investigated with a particular focus of scaling the resonator to the spatio-temporal characteristics of the most intense near-wall vertical velocity and wall-pressure fluctuations. Investigations of two neck-orifice diameters of $d^+ = 68$ and $d^+ = 102$, both with three resonance frequencies relating to a spatial scale of $\lambda_x^+ \equiv U^+_c/f^+ \approx 250$ (the inner-spectral peak of $v$ and $p_w$) and sub- and super-wavelength scaling tuned to $\lambda_x^+ \approx 126$ ($- 46$\,\%) and $\lambda_x^+ \approx 313$ ($+ 35$\,\%), respectively, resulted in several conclusions:\\[-14pt]
\begin{enumerate}[labelwidth=0.60cm,labelindent=0pt,leftmargin=1.00cm,label=(\roman*),align=left]
\item \noindent All HRs beneath the grazing flow behave acoustically: resonance frequencies were relatively well-predicted to within $\pm 10$\,\%, through classical expressions of acoustic resonators with end-corrections.\vspace{-6pt}
\item \noindent The smaller resonator with $d^+ = 68$ experienced much larger frictional losses (damping coefficient roughly 75\,\% to 100\,\% larger), so that its effect on the grazing TBL flow was minor: particularly in the large-scale range of turbulence scales where a negligible difference was observed, while the larger resonator with $d^+ = 102$ resulted in an energy attenuation.\vspace{-6pt}
\item \noindent For the larger neck diameter of $d^+ = 102$, non-time resolved PIV-based velocity fields revealed that perturbations in $v$ reached beyond $y^+ \approx 100$, evidencing a coupled response of the resonator with the grazing $v$ fluctuations of the near-wall cycle; $u$ perturbations of the resonance-scale were only noticeable in a region confined to $y^+ \lesssim 30$.\vspace{-6pt}
\item \noindent Persistence of the perturbations of the resonance-scale downstream of the resonator was rather limited: in all resonance cases the effects were not coherent beyond a convective distance of two resonance periods.\vspace{-6pt}
\item \noindent Spectral analyses of time-resolved hot-wire data (along a wall-normal profile $135l^*$ downstream of the neck-trailing edge) revealed that in the near-wall region, large-scale energy (at a temporal frequency roughly one decade smaller than the resonance frequency) was attenuated by over 20\,\%.\vspace{-6pt}
\item \noindent Conditional averaging of the non-time resolved velocity fields, to the resonance cycle of the cavity pressure, exposed that at resonance the low-frequency resonator comprised a phase relation between the vertical velocity at the neck-inlet, and the co-existing pressure, according to a classical resonator at resonance (in-phase behavior). A drift in this phase relation was observed for the higher-frequency resonator, which is ascribed to an effective smaller diameter when the stroke ratio becomes smaller than unity.
\end{enumerate}

Future impact of this work will be in the development of single configuration HRs and multi-interconnected ones, to attempt a stronger attenuation of (large-scale) energy in the flow. It is conjectured that large-scale energy attenuation can alter the near-wall Reynolds stress and the mean skin friction generation \citep{renard:2016a}.

\begin{acknowledgments}
Financial support of the European Office of Aerospace Research \& Development (EOARD) of the U.S. Air Force Office of Scientific Research (AFOSR) under Award No. FA8655-22-1-7168 (Intnl. Program  Officer Dr. Douglas Smith \& Program Officer Dr. Gregg Abate) is gratefully acknowledged. We also wish to acknowledge the Department of Flow Physics \& Technology of the Faculty of Aerospace Engineering at the Delft University of Technology, for financial support in establishing the experimental setup. We like to give special thanks to Stefan Bernardy, Peter Duyndam and Frits Donker Duyvis for technical assistance.
\end{acknowledgments}

\bibliography{bibtex_database}

\end{document}